\documentclass[conference]{IEEEtran}
\IEEEoverridecommandlockouts

\usepackage{cite}
\usepackage{amsmath,amssymb,amsfonts}
\usepackage{algorithm}
\usepackage{multirow}
\usepackage{longtable}
\usepackage{tabularx,booktabs}
\usepackage[table,xcdraw]{xcolor}
\usepackage{algpseudocode}
\usepackage{graphicx}
\usepackage{adjustbox}
\usepackage{float}
\usepackage{textcomp}
\usepackage{xcolor}
\def\BibTeX{{\rm B\kern-.05em{\sc i\kern-.025em b}\kern-.08em
    T\kern-.1667em\lower.7ex\hbox{E}\kern-.125emX}}
\begin{document}
\title{Comparison study of the combination of the SPSA algorithm and the PSO algorithm \\
}
\author{\IEEEauthorblockN{Bertrand Ngansop, Stefan Götz, Martin Eckl}
\IEEEauthorblockA{%\textit{Mat.Nr: 403381} \\
\textit{Department of Electrical and Computer Engineering}\\TU Kaiserslautern \\
ngansop@rhrk.uni-kl.de}
}

\maketitle

\begin{abstract}
Particle swarm optimization (PSO) is attracting an ever-growing attention and more than ever it has found many application areas for many challenging optimization problems.
It is, however, a known fact that PSO has a severe drawback in the update of its global best (gbest) particle, which has a crucial role of guiding the rest of the swarm. In this paper, we propose three efficient solutions to remedy this problem using the SPSA Algorithm. In the first approach, gbest is updated with respect to a global estimation of the gradient and can avoid getting trapped into a local optimum. The second approach is
based on the formation of an alternative or artificial global best particle, the so-called aGB, which can replace the native gbest particle for a better guidance, the decision of which is held by a fair competition between the two. The third approach is based on the update of the swarm particle. For this purpose we use simultaneous perturbation stochastic approximation (SPSA) for its low cost. Since SPSA is applied only to the gbest (not to the entire swarm) or to the entire swarm, both approaches result thus in a negligible overhead cost for the entire PSO process. Both approaches are shown to significantly improve the performance of PSO over a wide range of non-linear functions, especially if SPSA and PSO parameters are well selected to fit the problem at hand. As in the basic PSO application, experimental results show that the proposed approaches significantly improved the quality of the Optimization process as measured by a statistic analysis.
\end{abstract}

\section{Introduction}
Particle swarm optimisation (PSO) was introduced by Kennedy and Eberhart \cite{488968} in 1995 as a population-based stochastic search and optimisation method. PSO is widely used in science and engineering, particularly to optimise highly nonlinear multidimensional problems and often even physical plants, such as technical devices, electromagnetics, chemical processes, batteries, or even the body \cite{1282114,OURIQUE20021783,PRATA20093953,TANG20091391,9254880, 9589293,6347004,10.1371/journal.pone.0055771}. It arose from computer simulation of the individuals (particles or living organisms) in a flock of birds \cite{Barker1975SociobiologyTN}, which basically exhibit natural behaviour when searching for a target (e.g., food). Therefore, the algorithm has certain similarities with the other evolutionary algorithms (EAs) \cite{Bck1993AnOO} such as the genetic algorithm (GA) \cite{Goldberg1988GeneticAI}, genetic programming (GP) \cite{Koza1993GeneticP}, evolutionary strategies (ES) \cite{Bck1995EVOLUTIONARYAF}, and evolutionary programming (EP) \cite{Zaki2010AdvancesIK}. Their common ground is that evolutionary algorithms are population-based and they can avoid being trapped in a local optimum. So they can find optimal solutions. However, it is never guaranteed.
A major drawback of the algorithm lies in the direct connection of the information flow between particles and the global best particle $gbest$, which then guides the rest of the swarm. This leads to the formation of similar particles with some loss of diversity. This phenomenon thus increases the probability of being trapped in local optima \cite{Riget2002ADP}, and it is the main cause of the problem of premature convergence, especially when the search space has high dimensions \cite{Bergh2002AnAO} and the problem to be optimised is multimodal \cite{Riget2002ADP}. 

Moreover, PSO is an optimisation method that is sensitive to perturbations. In PSO, particles explore the search space by iteratively passing on the best solutions they find and their respective quality to other particles in the swarm. These interactions allow them to develop new solutions based on their own experience and that of others. However, the quality of such solutions depends largely on how accurate the information about the problem is. That is, if the information is noisy, the particles will be driven to solutions whose quality may be significantly worse than expected. Therefore, the performance of PSO is degraded in noisy environments.

An approach that we consider more reliable to deal with local convexity and noise in large optimisation problems, which is omnipresent when measurement data are involved in physical plants, and still low probability of being trapped in local optima combines particle swarm optimisation with stochastic approximation, such as simultaneous perturbation stochastic approximation (SPSA). Due to efficient gradient approximation under noise and variability influence, the SPSA algorithm is suitable for high-dimensional problems where many terms are determined in the optimisation process. Furthermore, it allows the objective function to consist of noisy measurements. The algorithm has desirable properties for both global and local optimisation in the sense that the gradient approximation is sufficiently noisy to avoid local minima, and at the same time sufficiently informative about the slope of the function to facilitate global convergence. To this end, in this paper we will analyse three approaches to address the drawbacks of particle swarm optimisation in an efficient and generic manner. The different approaches will be proposed in the next steps.

\section{Particle swarm optimisation}\label{pso}

Particle Swarm optimisation (PSO) is a popular meta-heuristic based on the social interaction of individuals living together in groups and cooperating with each other. It has attracted attention from a growing number of researchers because of its simplicity and efficiency \cite{Helwig2010ParticleSF}, \cite{Seo2006MultimodalFO}, \cite{Chen2010AnIC}. 

The goal of the PSO algorithm is to find the optimum of an objective function $$ f : S \subset R^n \rightarrow R$$. 
\begin{figure}[H]
 \centering
 \includegraphics[width=0.4\textwidth,angle=0]{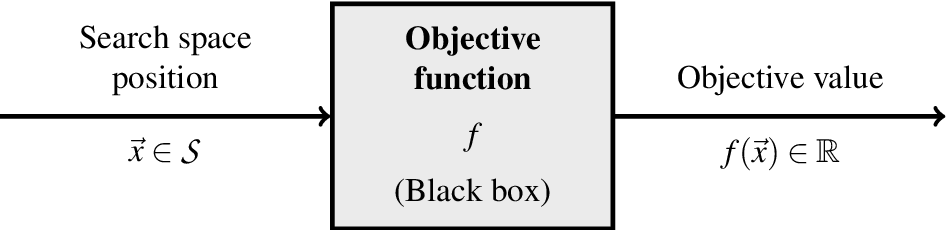}
 \caption[Black-Box]{Black-box optimisation \cite{Helwig2010ParticleSF}}
\label{fig:Black-Box}
\end{figure}
For the rest of this paper, we assume that $f$ is to be minimised. Particle swarm optimisation is often used in practice when there is no closed form of $f$. In such a situation, information about $f$ can only be obtained by evaluating $f$ pointwise. In particular, the information about the slope of $f$ is not available, which is usually relevant for finding the optimum of the target variable. Figure gives a graphical overview of the described situation, which is called a black-box optimisation problem (see Figure \ref{fig:Black-Box}).

\subsection{The PSO algorithm}\label{pso-algorithm}

The first version of a particle swarm optimisation algorithm was published by Kennedy and Eberhart \cite{488968}. The algorithm was designed to simulate a population of individuals, such as flocks of birds or schools of fish, searching for a region that is optimal with respect to some objective function, such as the quantity and quality of food. Unlike other popular nature-inspired meta-heuristics such as evolutionary algorithms (EAs), the particles of a particle swarm work together and exchange information about good places, rather than against each other.

The PSO algorithm takes into account two main sources of influence for social learning processes: Individuals rely on their own past experiences (cognitive component), and they imitate better group members (social component). They are implemented in an iteration-based optimisation algorithm as follows:

A population of $p$ individuals, hereafter called particles, explore the n-dimensional search space $S$ of an optimisation problem with objective function $f : S \subseteq R^n \rightarrow R$. Each particle $i$ has a position $\vec{x}_{i,k}$ (where $k$ is the iteration counter), a fitness value $f(\vec{x}_{i,k})$, and moves through the search space at a speed $\vec{v}_{i,k}$. The best search space position that particle $i$ has visited by iteration $k$ is called its personal best position $\vec{p}_{i,k}$. Each particle is assigned a subset of all particles as its neighbourhood. The best position visited by all particles up to iteration $k$ is called the global best position $\vec{G}_{k}$. In addition to the cognitive and social components, and following the model of flocks of birds or schools of fish, each particle additionally retains some of its old velocity, resulting in the following update equations for swarm optimisation:
\begin{multline}
\vec{v}_{i,k} = \omega \cdot \vec{v}_{i,k-1} + \underbrace {c1 \cdot \vec{r}_{1,i,k} \cdot ( \vec{p}_{i,k-1} - \vec{x}_{i,k-1})}_{cognitive \, component} \\ + \underbrace {c2 \cdot \vec{r}_{2,i,k} \cdot ( \vec{G}_{k-1} - \vec{x}_{i,k-1})}_{social \, component}
\label{update-equation}
\end{multline}
\begin{equation}
\vec{x}_{i,k} = \vec{x}_{i,k-1}+ \vec{v}_{i,k}
\label{iteration-equation}
\end{equation}
where $\omega$, $c_1$, and $c_2$ are given parameters, $\vec{r}_{1,i,k}$ and $ \vec{r}_{2,i,k}$ are vectors of real random numbers whose components are uniformly distributed in the interval $[0,1]$.
\begin{figure}[H]
 \centering
 \includegraphics[width=0.35\textwidth,angle=0]{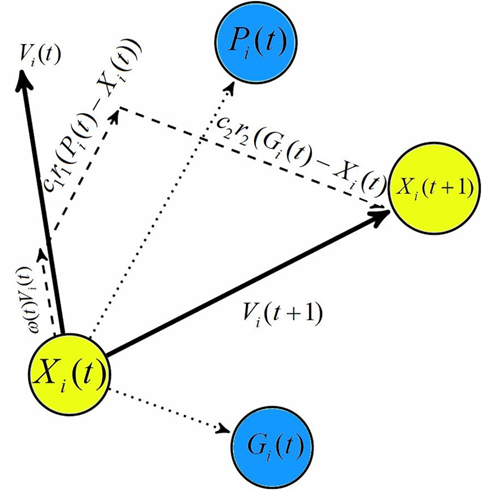}
 \caption[particle-component]{Component of a particle $i$ in the swarm}
\label{fig:particle-component}
\end{figure} 

After each particle has calculated its new position and velocity (see Figure \ref{fig:particle-component}), the personal best position of all particles are updated. Normally, the personal best position of a particle $\vec{p}_{i,k}$ is updated to its current position $\vec{x}_{i,k}$ if $f(\vec{x}_{i,k})$ $<$ $f(\vec{p}_{i,k})$ holds. To update the private guides, the objective function $f$ must be evaluated once for each particle in each iteration. The PSO is shown in Algorithm \ref{alg:bpso}. It is terminated as soon as a certain termination criterion is met, e.g., as soon as the best solution found has not further improved or the iteration counter k exceeds a certain limit. When applying particle swarm optimisation to a specific problem, the parameters of the algorithm, e.g., $\omega$, $c_1$, $c_2$, must be chosen accordingly.

\subsection{Initialisation of parameters}\label{parameters_initialization}

We will now discuss the initialisation of the particle swarm optimisation parameters. The inertia weight $\omega$ was first introduced by Shi and Eberhart in 1998 \cite{Eberhart2000ComparingIW}. It is mostly used smaller than 1 to reduce the exploration behaviour of the swarm over time. The so-called acceleration coefficients or control parameters $c_1$ and $c_2$ determine the relative influence of the cognitive and social components on the movement of a particle. The population size $p$ is often set to values between 20 and 50 and should be chosen problem characteristics and dimensionality. It appears that the size of the population only slightly affects the performance of the particle swarm.

Based on a deterministic PSO model, Clerc and Kennedy \cite{Clerc2002ThePS} showed that particle velocity limitation is not necessary to obtain a convergent particle swarm. Nevertheless, the use of velocity limiting can significantly improve the performance of a PSO algorithm. A detailed discussion of velocity limiting, including time-dependent and adaptive settings was presented by Engelbrecht \cite{Zyl2015ASM}. By adjusting the parameters, the trade-off between searching in areas not yet visited, and refining already good search points, can be influenced. Initialisation of positions is usually done uniformly at random over a limited search space. An alternative is presented in \cite{Richards2004ChoosingAS}, where the authors propose a method based on centroidal Voronoi mosaics to ensure that particles are distributed more uniformly over the search space than with a purely random distribution. The velocities of the particles can be initialised using one of the following ways:
\begin{itemize}
\item Uniform: The particle velocities are drawn uniformly at random in a given n-dimensional space;
\item Zero: The velocities are initialised to zero \cite{Engelbrecht2005FundamentalsOC}.
\end{itemize}

In a PSO process and for iteration $k$, each particle $i$ in the swarm $\xi=\{x_1, ..., x_a, ..., x_S\}$, is represented by the following properties:
\begin{itemize}
\item $xx^{xd_i(k)}_{i,j} (k)$ : $j$-th component (dimension) of the velocity of particle i, in dimension $xd_i(k)$.
\item $vx^{xd_i(k)}_{i,j} (k)$ : $j$-th component (dimension) of the velocity of particle i, in dimension $xd_i(k)$
\item $xy^{xd_i(t)}_{i,j} (k)$ : $j$-th component (dimension) of the personal best (\textit{pbest}) position of particle $i$, in dimension $xd_i(k)$
\item $\vec{G}(d)$: Global best particle index in dimension d
\item $xy^d_j (t)$ : $j$-th component of the global best position of the swarm, in dimension $d$
\item $xd_i(t)$: dimension component of particle $i$
\item $vd_i(t)$: velocity component of the dimension of particle $i$
\item $\vec{xd_i}(t)$ : personal best dimension component of particle $i$
\end{itemize}

\begin{algorithm}
\caption{bPSO (termination criteria: \textit{IterNo}, cut-off error = $10^{-4}$)}\label{alg:bpso}
\begin{algorithmic}[1]
\ForAll {$i \in [1, S]$}
	\State Randomize $xd_i(0)$ , $vd_i(0) = 0$
	\State Initialize $\vec{xd_i}(0) = xd_i(0)$
\EndFor
\ForAll {$k \in [1, IterNo]$}
	\ForAll {$i \in [1, S]$}
		\State Compute $pbest$
		\If {$(f(p_i(k)) < \min (f(\hat{p}(k-1)),\underset{1\leq j<i}{f(p_j (t))})$}
			\State $gbest = i \,\textrm{and} \, \hat{p}(t) = p_i(t)$
		\EndIf
	\EndFor
	\State If any termination criterion is met, then Stop.
	\ForAll {$i \in [1, S]$}
		\ForAll {$j \in [1, N]$}
			\State Compute $v_{i,j}(t+1)$ using Eq.\  \ref{update-equation}
			\If {$v_{i,j}(t+1) > V_{max}$}
				\State $V_{i,j}(t+1) = V_{max}$
			\EndIf
			\State Compute $4$ using Eq.\ \ref{iteration-equation}
		\EndFor
	\EndFor
\EndFor
\end{algorithmic}
\end{algorithm}

\subsection{Simultaneous Perturbation Stochastic Approximation}\label{SPSA}
The simultaneous perturbation stochastic approximation (SPSA) method was introduced by Spall \cite{Spall1998ImplementationOT}. It is based on an easy-to-implement and highly efficient gradient approximation based on the measurement of the objective function rather than the measurement of the gradient of the objective function. Here, only two measurements of the objective function are needed, regardless of the number of parameters to be optimised, to estimate the derivative at each run. This contrasts with the Kiefer-Wolfowitz algorithm in that $2p$ measurements are required, where $p$ is the number of parameters to be optimised, to approximate the gradient. Furthermore, it was proved by Spall that the SPSA achieves the same statistical accuracy as FDSA for a given number of runs, although SPSA uses p times fewer function evaluations than FDSA. The objective of the SPSA algorithm is to solve the problem
\begin{equation}
\mathop{\min}_{\theta} L(\theta)
\label{target_spsa}
\end{equation}
where $L$ represents the objective function. Here it is assumed that there are only noisy measurements $y(\theta) = L(\theta) + \varepsilon(\theta)$ ($\varepsilon(\theta)$ represents the noise term) of the objective function and L is a differentiable function of $\theta$.  Here, the SPSA procedure uses the same iterative process as the Kiefer-Wolfowitz procedure per
\begin{equation}
\hat \theta_{k+1} = \hat \theta_k - a_k \hat g_k (\hat \theta_k),
\label{setup_spsa}
\end{equation}
Where $a_k$ is a positive non-random sequence that approaches zero as $k$ grows to infinity, and $\hat g_k$ is the simultaneous perturbation approximation to the unknown derivative $g(\hat \theta_k) = \frac{\partial L(\hat \theta_k)}{\partial \hat \theta_k}$. $\hat g_k$ is calculated as
\begin{equation}
\hat g_k(\hat \theta_k) = \frac{y(\hat \theta_k) + c_k \Delta_k - y(\hat \theta_k) - c_k \Delta_k}{2c_k}
\begin{bmatrix}
\Delta_{k1}^{-1} \\
\Delta_{k2}^{-1} \\
\vdots \\
\Delta_{kp}^{-1}
\end{bmatrix},
\end{equation}
Where $c_k$ is again a positive non-random sequence that approaches zero as $k$ grows to infinity, and $\Delta_{ki}$ is the $i$-th component of the vector $\Delta_k$ representing the distribution of perturbation terms. To ensure the efficiency of the algorithm, the parameters of the algorithm should satisfy the conditions
\begin{equation}
\sum_{k=1}^{\infty} a_k = \infty,
\end{equation}
\begin{equation}
c_k \rightarrow 0 \, \text{for} \, k \rightarrow \infty,
\end{equation}
\begin{equation}
\sum_{k=1}^{\infty} a_k^2 c_k^{-2} < \infty.
\end{equation}

$\Delta_{ki}$ must be independent random variables with zero mean, symmetrically distributed around zero.

It should be noted that the algorithm was designed for unconstrained problems. However, by applying penalty methods, it is possible to use SPSA on a constrained problem.  

\section{The proposed technique}

\subsection{First SPSA--PSO approach}\label{gbest_update}

In the PSO procedure, a swarm of particles (or agents), each representing a potential solution to an optimisation problem, navigates the search space (or solution space). At each time $t$, every particle $i$ has a current position $\vec{x}_{i,k}$ and a velocity $\vec{v}_{i,k}$. In addition, each particle remembers the best position it has visited so far. This position is called the local attractor or private guide and is denoted by $\vec{p}_{i,k}$. The best of all the local attractors in the swarm is called the global attractor. This particular position is denoted $\vec{G}_{k}$ and it is visible to every particle. By updating the global attractor, a particle shares its information with the rest of the swarm. This clarifies that in each iteration of a PSO procedure $\vec{G}_{k}$ is the most important particle. However, it has the worst updating equation, i.e., when a particle becomes $\vec{G}_{k}$, it is at its personal best position ($\vec{p}_{i,k}$) and thus both social and cognitive components in the velocity updating equation (see Eq.\ \ref{update-equation}) are cancelled. Thus, if $\vec{G}_{k}$ is trapped in a local optimum, the rest of the swarm risks being trapped in a local optimum. The idea of the first approach simply improves and updates the global best position after each iteration using the SPSA algorithm \cite{Kiranyaz2009StochasticAD}. In that case, $\vec{G}_{k}$ is chosen as the initial parameter for the SPSA procedure. The SPSA exploits local convexity of the average cost function and pushes the global best into the nearby average minimum. The resulting $\vec{G'}_{k}$ then replaces the old global best position $\vec{G}_{k}$ of the swarm. The pseudo-code of the approach based on the algorithm first proposed in \cite{Kiranyaz2009StochasticAD} is shown belown.
%\marginpar{We should potentially put the algorithms into floating environments so that they can be positioned as flexible as figures.}
\begin{algorithm}[ht]   %[H]
\caption{SA-PSO (1) (termination criteria: \textit{IterNo}, cut-off error = $10^{-4}$, \textit{S, a, c, A, $\alpha$, $\gamma$})}\label{alg:sa-pso(1)}
\begin{algorithmic}[1]
\ForAll {$i \in [1, S]$}
	\State Randomize $xd_i(0)$ , $vd_i(0) = 0$
	\State Initialize $\vec{xd_i}(0) = xd_i(0)$
\EndFor
\ForAll {$k \in [1, IterNo]$}
	\ForAll {$i \in [1, S]$}
		\State Compute $pbest$
		\If {$(f(p_i(k)) < \min (f(\hat{p}(k-1)),\underset{1\leq j<i}{f(p_j (t))})$}
			\State $gbest = i \,\textrm{and} \, \hat{p}(t) = p_i(t)$
		\EndIf
	\EndFor
	\State If any termination criterion is met, then Stop.
	\ForAll {$i \in [1, S]$}
		\ForAll {$j \in [1, N]$}
			\State Compute $v_{i,j}(t+1)$ using Eq.\  \ref{update-equation}
			\If {$v_{i,j}(t+1) > V_{max}$}
				\State $V_{i,j}(t+1) = V_{max}$
			\EndIf
			\State Compute $4$ using Eq.\ \ref{iteration-equation}
			\If {$i = gbest$}
				\ForAll {$k \in [1, IterNo]$}
					\State Initialize $\hat{\theta_0} = gbest$
					\State Compute $a_k$ and $c_k$ 
					\State Compute $L(\hat \theta_k + c_k \Delta_k)$ and $L(\hat \theta_k - c_k \Delta_k)$
					\State Compute $\hat{g}_k(\hat \theta_k)$ 
					\State Compute $\hat \theta_{k+1}$
				\EndFor
			\State Compute $\hat \theta_{k+1} = gbest$
			\EndIf
		\EndFor
	\EndFor
\EndFor

\end{algorithmic}
\end{algorithm}

\subsection{Second SPSA--PSO approach}\label{gbest}
%\marginpar{is GB global best? Spell out every (!) acronym before the first use!}
The second approach has a similar motivation to the fractional global best (GB) formation (FGBF) proposed in \cite{Kiranyaz2010FractionalPS}. FGBF \cite{Kiranyaz2010FractionalPS} was designed to avoid premature convergence by providing significant diversity through proper fusion of the best components of the swarm. At each iteration in a PSO process, an artificial GB (aGB) particle is formed (fractionated) by taking the most promising (or simply the best) particles
%\marginpar{How many best particles? The M best or only one?}
from the entire swarm. Therefore, especially in the first steps, FGBF may be a better alternative than the $gbest$ particle optimised from the particle swarm. This process naturally exploits the available diversity of each dimensional component and thus can prevent the swarm from being trapped in local optima. Therefore, in the proposed method, the best components from each particle are collected to create an artificial GB candidate, namely aGB, which will replace the global best position $\vec{G}_{k}$ particle of the swarm if it is better than the previous global best particle. It should be noted here that whenever a better (real) $gbest$ particle or aGB particle appears, it will replace the current global best particle. Therefore, without using any of the above modifications, we will show that the proposed fractional PSO can avoid local optima and thus find the optimum (or close to optimum) even in high-dimensional search spaces and usually at earlier stages. In consequence, aGB is only used if it is better than the global best and replaces the same; otherwise, the results are ignored and forgotten. The pseudo-code of the approach based on the algorithm first proposed in \cite{Kiranyaz2009StochasticAD} is presented below.
\begin{algorithm}[ht]   %[H]
\caption{SA-PSO (2) (termination criteria: \textit{IterNo}, cut-off error = $10^{-4}$, \textit{S, a, c, A, $\alpha$, $\gamma$})}\label{alg:sa-pso(2)}
\begin{algorithmic}[1]
\ForAll {$i \in [1, S]$}
	\State Randomize $xd_i(0)$ , $vd_i(0) = 0$
	\State Initialize $\vec{xd_i}(0) = xd_i(0)$
\EndFor
\ForAll {$k \in [1, IterNo]$}
	\ForAll {$i \in [1, S]$}
		\State Compute $pbest$
		\If {$(f(p_i(k)) < \min (f(\hat{p}(k-1)),\underset{1\leq j<i}{f(p_j (t))})$}
			\State $gbest = i \,\textrm{and} \, \hat{p}(t) = p_i(t)$
		\EndIf
	\EndFor
	\State If any termination criterion is met, then Stop.
	\ForAll {$i \in [1, S]$}
		\ForAll {$j \in [1, N]$}
			\State Compute $v_{i,j}(t+1)$ using Eq.\  \ref{update-equation}
			\If {$v_{i,j}(t+1) > V_{max}$}
				\State $V_{i,j}(t+1) = V_{max}$
			\EndIf
			\State Compute $4$ using Eq.\ \ref{iteration-equation}
			\If {$i = gbest$}
				\ForAll {$k \in [1, IterNo]$}
					\State Create a new aGB particle, $\{(xx^{d}_{aGB}(t+1), xy^{d}_{aGB}(t+1))\} \forall d \in [d_{min}, D_{max}]$
					\State Let $\hat{\theta_0} = aGB$
					\State Compute $a_k$ and $c_k$ 
					\State Compute $L(\hat \theta_k + c_k \Delta_k)$ and $L(\hat \theta_k - c_k \Delta_k)$
					\State Compute $\hat{g}_k(\hat \theta_k)$ 
					\State Compute $\hat \theta_{k+1}$
				\EndFor
				\If {$f(xy^{d}_{aGB}(t+1)) < f(xy^{d}_{gbest}(t))$}
					\State $xy^{d}_{gbest}(t) = xy^{d}_{aGB}(t+1)$
				\EndIf
			\EndIf
		\EndFor
	\EndFor
\EndFor

\end{algorithmic}
\end{algorithm}

\subsection{Third SPSA--PSO approach}\label{pbest_update}

In PSO, particles explore the search space by iteratively passing the best solutions they find and their respective quality to other particles in the swarm. These interactions allow them to develop new solutions based on their own \textit{experience} or history and that of others. With noisy measurements of the objective function, there is then a high probability that the particles will be driven to solutions whose quality may be significantly worse than expected. In this approach, each particle is then updated using the SPSA procedure, as it is much more robust than particle swarm optimisation. That is, if a particle changes $\vec{G}_{k}$ after one iteration using the update equation, it is updated, i.e., moved to the new position, using the SPSA algorithm before the next iteration. The pseudo-code of the above algorithm is shown below.
\begin{figure}[htb!]
 \centering
 \includegraphics[width=0.4\textwidth,angle=0]{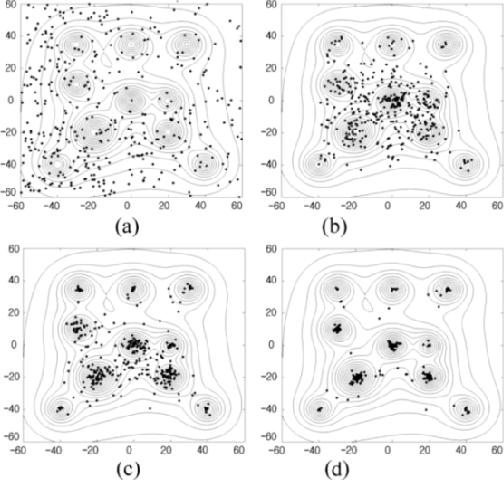}
 \caption[multi-pso]{SPSA--PSO (3) over a test function \cite{Seo2006MultimodalFO}}
\label{fig:multi-pso}
\end{figure}

\begin{algorithm}[hbt!]
\caption{SA-PSO (3) (termination criteria: \textit{IterNo}, cut-off error = $10^{-4}$, \textit{S, a, c, A, $\alpha$, $\gamma$})}\label{alg:sa-pso(3)}
\begin{algorithmic}[1]
\ForAll {$i \in [1, S]$}
	\State Randomize $xd_i(0)$ , $vd_i(0) = 0$
	\State Initialize $\vec{xd_i}(0) = xd_i(0)$
\EndFor
\ForAll {$k \in [1, IterNo]$}
	\ForAll {$i \in [1, S]$}
		\State Compute $pbest$
		\If {$(f(p_i(k)) < \min (f(\hat{p}(k-1)),\underset{1\leq j<i}{f(p_j (t))})$}
			\State $gbest = i \,\textrm{and} \, \hat{p}(t) = p_i(t)$
		\EndIf
	\EndFor
	\State If any termination criterion is met, then Stop.
	\ForAll {$i \in [1, S]$}
		\ForAll {$j \in [1, N]$}
			\State Compute $v_{i,j}(t+1)$ using Eq.\  \ref{update-equation}
			\If {$v_{i,j}(t+1) > V_{max}$}
				\State $V_{i,j}(t+1) = V_{max}$
			\EndIf
			\State Compute $4$ using Eq.\ \ref{iteration-equation}
			\If {$i = gbest$}
				\ForAll {$k \in [1, IterNo]$}
					\State Initialize $\hat{\theta_0} = gbest$
					\State Compute $a_k$ and $c_k$ 
					\State Compute $L(\hat \theta_k + c_k \Delta_k)$ and $L(\hat \theta_k - c_k \Delta_k)$
					\State Compute $\hat{g}_k(\hat \theta_k)$ 
					\State Compute $\hat \theta_{k+1}$
				\EndFor
			\State Compute $\hat \theta_{k} = gbest$
			\EndIf
		\EndFor
	\EndFor
\EndFor

\end{algorithmic}
\end{algorithm}

\section{Results of the Numerical Analysis}\label{results_analysis}

\begin{figure}[htb!]
 \centering
 \includegraphics[width=0.4\textwidth,angle=0]{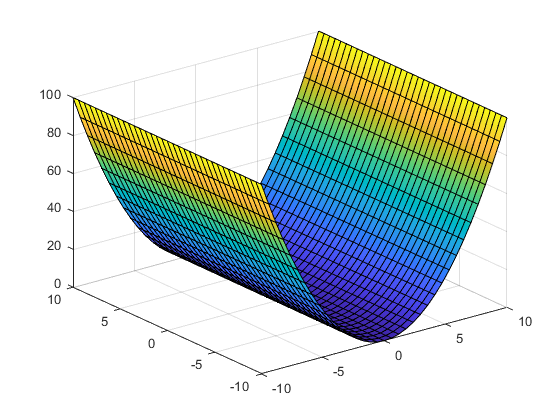}
 \caption[sphere]{Sphere in 2-D}
\label{fig:sphere}
\end{figure}

\begin{figure}[htb!]
 \centering
 \includegraphics[width=0.4\textwidth,angle=0]{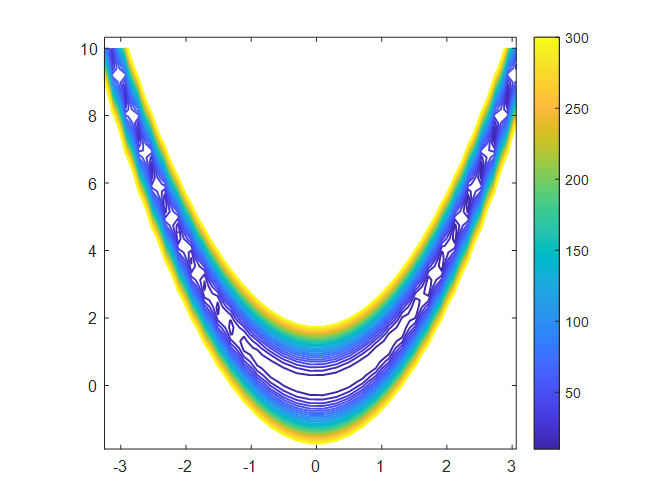}
 \caption[rosenbrock]{Rosenbrock in 2-D}
\label{fig:rosenbrock}
\end{figure}

\begin{figure}[htb!]
 \centering
 \includegraphics[width=0.4\textwidth,angle=0]{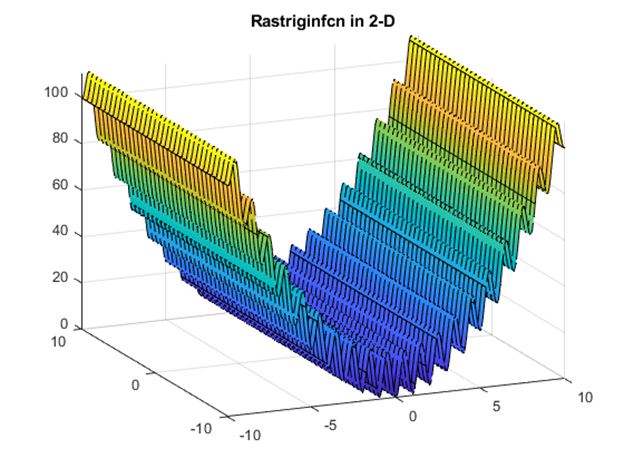}
 \caption[rastrigin]{Rastrigin in 2-D}
\label{fig:rastrigin}
\end{figure}

Here, we test and evaluate both approaches of the SPSA--PSO in comparison to the basic PSO (bPSO) (see Algorithm \ref{alg:bpso}) and SPSA over several uni- and multimodal benchmark functions in high dimensions. We used three benchmark functions listed in Table \ref{tab:testfunction} to provide a good mix of complexity and modality and have been investigated by several researchers, see for example. \cite{Angeline1998EvolutionaryOV}, \cite{Esquivel2003OnTU} and \cite{Shi1998AMP}. The $Sphere$ function (see Figure \ref{fig:sphere}) and the $Rosenbrock$ function (see Figure \ref{fig:rosenbrock}) are the uni-modal functions and the $Rastrigin$ function (see Figure \ref{fig:rastrigin}) is multimodal, which means that it has many deceptive local minima. We used the same termination criteria as the maximum number of iterations allowed (MaxIter = 10000). Also, three dimensions (20, 50, and 80) are used for the example functions to test the performance of each method in these different dimensions individually. For the particle swarm optimisation and the three different approaches of the SPSA--PSO combination, we used the swarm size $S=50$.

\begin{table*}
\centering
\caption{testfunctions}
\begin{adjustbox}{max width=\textwidth}
\begin{tabular}{|l
>{\columncolor[HTML]{656565}}l l
>{\columncolor[HTML]{656565}}l |l}
\cline{1-4}
\textbf{Function} & \textbf{Formula} & \textbf{Range} & \textbf{Dimension} & \\ \cline{1-4}
\textbf{Sphere} & $F_1(x,d) = (\sum_{i=1}^d x_i^2)$ & $[-150, 150]$ & $20,\, 50, \, 80$ & \\  \textbf{Rosenbrock} & $F_2(x,d) = (\sum_{i=1}^d 100(x_{i+1}-x_i^2)^2 + (x_i -1)^2)$ & $[-50,25]$ & $20,\, 50, \, 80$ & \\ \textbf{rastrigin} & $F_3(x,d) = (\sum_{i=1}^d 10 +x_i^2-10\cos(2\pi x_i))$ & $[-150,150]$ & $20,\, 50, \, 80$ & \\ \cline{1-4}
\end{tabular}
\label{tab:testfunction}
\end{adjustbox}
\end{table*}


In this work, the parameters of the SPSA--PSO algorithm were also optimised. For this purpose, the results with the original parameters according to Kennedy and Eberhart \cite{Shi1998AMP} are compared with optimised parameters. The particle swarm optimisation parameters for the experiments in this paper are $c1 = c2 = 2$. We use these parameters for each experiment except for the experiments where the parameters were optimised.

By tuning the parameters, it is possible to balance the trade-off between exploration (the ability to search in areas not yet visited) and exploitation (the ability to refine search points that are already good). Clerc and Kennedy \cite{Clerc2002ThePS} analysed the PSO algorithm in both discrete and continuous time using a deterministic model similar to the one presented above (see chapter \ref{pso}). A so-called constricted particle swarm optimisation (PSO-chi) was developed by rewriting the original PSO updating equations to
\begin{multline}
\vec{v}_{i,k} = \chi \cdot (\vec{v}_{i,k-1} + \phi_1 \cdot \vec{x}_{1,i,k} \cdot (\vec{p}_{i,k-1} - \vec{x}_{i,k-1}) \\ + \phi_2 \cdot \vec{x}_{2,i,k} \cdot (\vec{G}_{i,k-1} - \vec{x}_{i,k-1})),
\end{multline}
\begin{equation}
\vec{x}_{i,k} = \vec{x}_{i,k-1} + \vec{v}_{i,k},
\end{equation}
Where $\chi$ is called the constriction coefficient. These updating equations are algebraically equivalent to the standard equations (see Eq.\ \ref{update-equation} and Eq. \ref{iteration-equation}) by $\chi = \omega$ and $\chi \cdot \phi_i = c_i$ for $i = 1,2$. Clerc and Kennedy proved that the dynamical system converges when the constriction coefficient $\chi$ is calculated as
\begin{equation}
\chi = \frac{2 \cdot \kappa}{\phi - 2 + \sqrt{\phi^2 - 4 \cdot \phi}},
\label{equ:chi}
\end{equation}
Where $\kappa \in [0,1]$ and $\phi = \phi_1 + \phi_2 > 4$. In our tests, we used the following parameters: $\kappa = 1$, $\phi_1 = 2.05$, $\phi_2 = 2.05$.

For the SPSA algorithm, we used the recommended values for $A$, $\alpha$, and $\gamma$: $200$, $0.602$ and $0.101$, which are fixed for all functions. We intentionally set the parameter setting for SPSA as this was feasible for these studies. To allow a fair comparison between SPSA, bPSO, and SPSA--PSO, the number of scores is kept the same.
\begin{figure}[htb]  %[H]
 \centering
 \includegraphics[width=0.4\textwidth,angle=0]{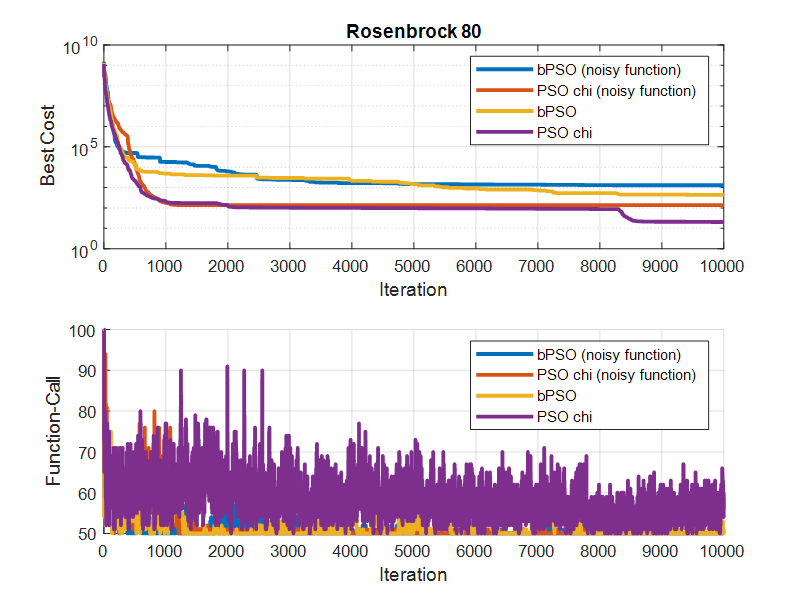}
 \caption[rosenbrock-bPSO]{bPSO on the Rosenbrock with dimension $d = 80$.}
\label{fig:rosenbrock-bpso}
\end{figure} 
\begin{figure}[htb] %[H]
 \centering
 \includegraphics[width=0.4\textwidth,angle=0]{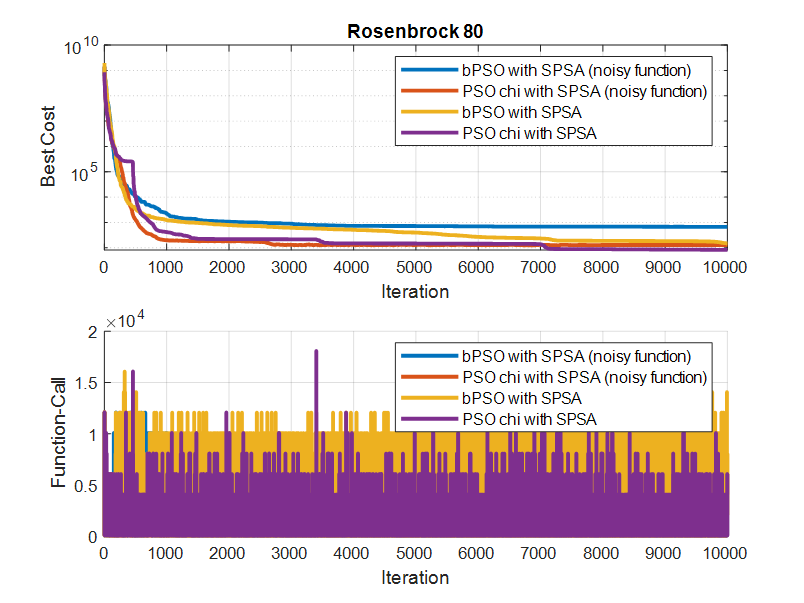}
 \caption[Rosenbrock-SPSA--PSO (1)]{SPSA--PSO (1) on the Rosenbrock with dimension $d = 80$.}
\label{fig:rosenbrock-spsapso-1}
\end{figure}
\begin{figure}[htb] %[H]
 \centering
 \includegraphics[width=0.4\textwidth,angle=0]{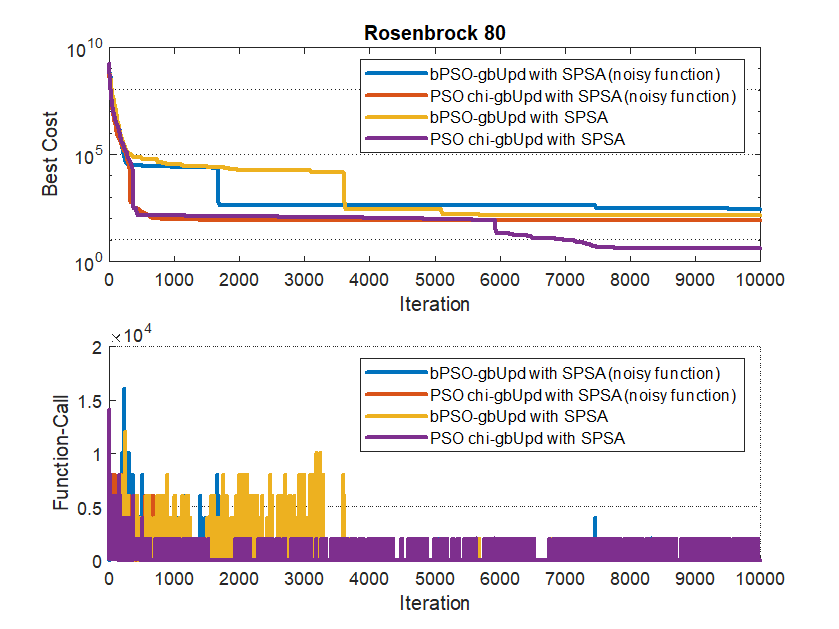}
\caption[Rosenbrock-SPSA--PSO (2)]{SPSA--PSO (2) on the Rosenbrock with dimension $d = 80$.}
\label{fig:rosenbrock-spsapso-2}
\end{figure}
\begin{figure}[htb] %[H]
 \centering
 \includegraphics[width=0.4\textwidth,angle=0]{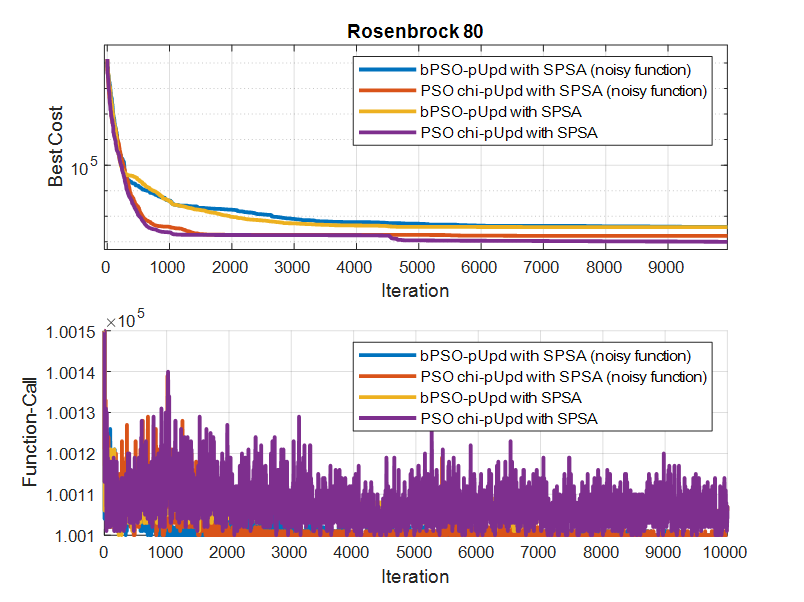}
 \caption[Rosenbrock-SPSA--PSO (3)]{SPSA--PSO(3) on the Rosenbrock with dimension $d = 80$.}
\label{fig:rosenbrock-spsapso-3}
\end{figure} 

Figures \ref{fig:rosenbrock-bpso}, \ref{fig:rosenbrock-spsapso-1}, \ref{fig:rosenbrock-spsapso-2}, and \ref{fig:rosenbrock-spsapso-3} show the courses of the optimisation with the algorithms to be compared. In addition to the progressions, the function calls per iteration are also shown. It can be seen that the three approaches proposed above achieve better results than the bPSO \ref{pso} even with noisy test functions. Moreover, it can be seen that adjusting the parameters (see Eq. \ref{equ:chi}) improves the PSO algorithm. Moreover, the third approach (see Eq. \ref{pbest_update}) has 100 times more function calls per iteration than the other two proposed approaches. Nevertheless, this approach achieves better results than the others. 

For each setting (for each function and dimension), 100 runs are performed and the first- and second-order statistics (mean $\mu$ and standard deviation $\sigma$) of the fitness scores are presented in Table \ref{statistik_100},  \ref{statistik_100-sphere}, and \ref{statistik_100-rastrigin}, with the best statistics highlighted. For each run, the process stops when the maximum number of iterations is reached. As the overall statistics on the right-hand side of Table \ref{statistik_100} show, both SPSA--PSO approaches outperform the $Rosenbrock$ function, regardless of dimension and modality and without exception. In other words, SPSA--PSO always performs better than the best bPSO. This basically confirms our claim above, i.e., the PSO update for $gbest$ is so poor that even adjusting the parameters of the PSO algorithm can still significantly improve the overall performance. 

\begin{table}[]
 \centering
\caption{Statistical results from 100 runs over the $Rosenbrock$-Function}
\begin{adjustbox}{max width=\paperwidth, angle=90}
\begin{tabular}{|c|c|cc|cc|cc|cc|cc|cc|cc|cc|}
\hline
 &  & \multicolumn{2}{c|}{\textbf{bPSO}} & \multicolumn{2}{c|}{\textbf{bPSO-chi}} & \multicolumn{2}{c|}{\textbf{SPSA--PSO (1)}} & \multicolumn{2}{c|}{\textbf{SPSA--PSO-chi (1)}} & \multicolumn{2}{c|}{\textbf{SPSA--PSO (2)}} & \multicolumn{2}{c|}{\textbf{SPSA--PSO-chi (2)}} & \multicolumn{2}{c|}{\textbf{SPSA--PSO (3)}} & \multicolumn{2}{c|}{\textbf{SPSA--PSO-chi (3)}} \\ \cline{3-18} 
\multirow{-2}{*}{\textbf{Functions}} & \multirow{-2}{*}{\textbf{d}} & \multicolumn{1}{c|}{\textbf{$\mu$}} & \textbf{$\sigma$} & \multicolumn{1}{c|}{\textbf{$\mu$}} & \textbf{$\sigma$} & \multicolumn{1}{c|}{\textbf{$\mu$}} & \textbf{$\sigma$} & \multicolumn{1}{c|}{\textbf{$\mu$}} & \textbf{$\sigma$} & \multicolumn{1}{c|}{\textbf{$\mu$}} & \textbf{$\sigma$} & \multicolumn{1}{c|}{\textbf{$\mu$}} & \textbf{$\sigma$} & \multicolumn{1}{c|}{\textbf{$\mu$}} & \textbf{$\sigma$} & \multicolumn{1}{c|}{\textbf{$\mu$}} & \textbf{$\sigma$} \\ \hline
 & 20 & \multicolumn{1}{c|}{8.15} & 3.99 & \multicolumn{1}{c|}{0.57} & 1.4 & \multicolumn{1}{c|}{1.2} & 0.46 & \multicolumn{1}{c|}{0.2} & 1.08 & \multicolumn{1}{c|}{2.64} & 2.17 & \multicolumn{1}{c|}{\textbf{0.31}} & \textbf{1.08} & \multicolumn{1}{c|}{0.2} & 1.4 & \multicolumn{1}{c|}{{\color[HTML]{FE0000} \textbf{0.1}}} & {\color[HTML]{FE0000} \textbf{1.1}} \\ \cline{2-18} 
 & 50 & \multicolumn{1}{c|}{59.2} & 27.7 & \multicolumn{1}{c|}{15.7} & 7.81 & \multicolumn{1}{c|}{15.9} & 5.21 & \multicolumn{1}{c|}{1.29} & 0.48 & \multicolumn{1}{c|}{32.5} & 13.9 & \multicolumn{1}{c|}{\textbf{0.91}} & \textbf{1.68} & \multicolumn{1}{c|}{3.84} & 5.9 & \multicolumn{1}{c|}{{\color[HTML]{FE0000} \textbf{0.68}}} & {\color[HTML]{FE0000} \textbf{1.8}} \\ \cline{2-18} 
\multirow{-3}{*}{\textbf{Rosenbrock}} & 80 & \multicolumn{1}{c|}{213} & 115 & \multicolumn{1}{c|}{64.2} & 30.6 & \multicolumn{1}{c|}{70.9} & 10.5 & \multicolumn{1}{c|}{28.15} & 5.16 & \multicolumn{1}{c|}{61.4} & 10.5 & \multicolumn{1}{c|}{\textbf{0.75}} & \textbf{1.57} & \multicolumn{1}{c|}{9.69} & 10.37 & \multicolumn{1}{c|}{{\color[HTML]{FE0000} \textbf{0.7}}} & {\color[HTML]{FE0000} \textbf{1.2}} \\ \hline
 & 20 & \multicolumn{1}{c|}{12.5} & 3.64 & \multicolumn{1}{c|}{12.3} & 2.62 & \multicolumn{1}{c|}{20.9} & 12.2 & \multicolumn{1}{c|}{12.3} & 2.67 & \multicolumn{1}{c|}{12.4} & 2.61 & \multicolumn{1}{c|}{\textbf{12.3}} & \textbf{1.93} & \multicolumn{1}{c|}{2.75} & 2.47 & \multicolumn{1}{c|}{{\color[HTML]{FE0000} \textbf{2.2}}} & {\color[HTML]{FE0000} \textbf{1.15}} \\ \cline{2-18} 
 & 50 & \multicolumn{1}{c|}{83.4} & 31.2 & \multicolumn{1}{c|}{59.2} & 14.5 & \multicolumn{1}{c|}{89.4} & 46.23 & \multicolumn{1}{c|}{57.9} & 18.9 & \multicolumn{1}{c|}{48.5} & 6.38 & \multicolumn{1}{c|}{\textbf{45.7}} & \textbf{8.46} & \multicolumn{1}{c|}{13.9} & 3.45 & \multicolumn{1}{c|}{{\color[HTML]{FE0000} \textbf{10.55}}} & {\color[HTML]{FE0000} \textbf{7.8}} \\ \cline{2-18} 
\multirow{-3}{*}{\textbf{Rosenbrock + noise}} & 80 & \multicolumn{1}{c|}{319} & 68 & \multicolumn{1}{c|}{169} & 55.6 & \multicolumn{1}{c|}{170} & 58.5 & \multicolumn{1}{c|}{112} & 58 & \multicolumn{1}{c|}{109} & 27.6 & \multicolumn{1}{c|}{\textbf{85.1}} & \textbf{20.5} & \multicolumn{1}{c|}{30.8} & 15.7 & \multicolumn{1}{c|}{{\color[HTML]{FE0000} \textbf{27.8}}} & {\color[HTML]{FE0000} \textbf{12.8}} \\ \hline
\end{tabular}
\end{adjustbox}
\label{statistik_100}
\end{table}

\begin{table}[]
 \centering
\caption{Statistical results from 100 runs over the $Rastrigin$-Function}
\begin{adjustbox}{max width=\paperwidth, angle=90}
\begin{tabular}{|c|c|cc|cc|cc|cc|cc|cc|cc|cc|}
\hline
 &  & \multicolumn{2}{c|}{\textbf{bPSO}} & \multicolumn{2}{c|}{\textbf{bPSO-chi}} & \multicolumn{2}{c|}{\textbf{SPSA--PSO (1)}} & \multicolumn{2}{c|}{\textbf{SPSA--PSO-chi (1)}} & \multicolumn{2}{c|}{\textbf{SPSA--PSO (2)}} & \multicolumn{2}{c|}{\textbf{SPSA--PSO-chi (2)}} & \multicolumn{2}{c|}{\textbf{SPSA--PSO (3)}} & \multicolumn{2}{c|}{\textbf{SPSA--PSO-chi (3)}} \\ \cline{3-18} 
\multirow{-2}{*}{\textbf{Functions}} & \multirow{-2}{*}{\textbf{d}} & \multicolumn{1}{c|}{\textbf{$\mu$}} & \textbf{$\sigma$} & \multicolumn{1}{c|}{\textbf{$\mu$}} & \textbf{$\sigma$} & \multicolumn{1}{c|}{\textbf{$\mu$}} & \textbf{$\sigma$} & \multicolumn{1}{c|}{\textbf{$\mu$}} & \textbf{$\sigma$} & \multicolumn{1}{c|}{\textbf{$\mu$}} & \textbf{$\sigma$} & \multicolumn{1}{c|}{\textbf{$\mu$}} & \textbf{$\sigma$} & \multicolumn{1}{c|}{\textbf{$\mu$}} & \textbf{$\sigma$} & \multicolumn{1}{c|}{\textbf{$\mu$}} & \textbf{$\sigma$} \\ \hline
 & 20 & \multicolumn{1}{c|}{52} & 9.67 & \multicolumn{1}{c|}{22} & 10.5 & \multicolumn{1}{c|}{43.8} & 7.4 & \multicolumn{1}{c|}{6.8} & 1.8 & \multicolumn{1}{c|}{0} & 0 & \multicolumn{1}{c|}{\textbf{0}} & \textbf{0} & \multicolumn{1}{c|}{0} & 0 & \multicolumn{1}{c|}{{\color[HTML]{FE0000} \textbf{0}}} & {\color[HTML]{FE0000} \textbf{0}} \\ \cline{2-18} 
 & 50 & \multicolumn{1}{c|}{62.6} & 28.7 & \multicolumn{1}{c|}{55.7} & 17.81 & \multicolumn{1}{c|}{8.4} & 2.1 & \multicolumn{1}{c|}{7.9} & 3.58 & \multicolumn{1}{c|}{0} & 0 & \multicolumn{1}{c|}{\textbf{0}} & \textbf{0} & \multicolumn{1}{c|}{0} & 0 & \multicolumn{1}{c|}{{\color[HTML]{FE0000} \textbf{0}}} & {\color[HTML]{FE0000} \textbf{0}} \\ \cline{2-18} 
\multirow{-3}{*}{\textbf{Rastrigin}} & 80 & \multicolumn{1}{c|}{300.9} & 85 & \multicolumn{1}{c|}{164.2} & 86.6 & \multicolumn{1}{c|}{43.8} & 7.4 & \multicolumn{1}{c|}{5.83} & 2.38 & \multicolumn{1}{c|}{0} & 0 & \multicolumn{1}{c|}{\textbf{0}} & \textbf{0} & \multicolumn{1}{c|}{0} & 0 & \multicolumn{1}{c|}{{\color[HTML]{FE0000} \textbf{0}}} & {\color[HTML]{FE0000} \textbf{0}} \\ \hline
 & 20 & \multicolumn{1}{c|}{52.5} & 8.79 & \multicolumn{1}{c|}{22.3} & 12.62 & \multicolumn{1}{c|}{20.9} & 12.2 & \multicolumn{1}{c|}{12.3} & 2.67 & \multicolumn{1}{c|}{0} & 0 & \multicolumn{1}{c|}{\textbf{0}} & \textbf{0} & \multicolumn{1}{c|}{0} & 0 & \multicolumn{1}{c|}{{\color[HTML]{FE0000} \textbf{0}}} & {\color[HTML]{FE0000} \textbf{0}} \\ \cline{2-18} 
 & 50 & \multicolumn{1}{c|}{83.4} & 31.2 & \multicolumn{1}{c|}{59.2} & 14.5 & \multicolumn{1}{c|}{27.4} & 14.8 & \multicolumn{1}{c|}{13.2} & 1.7 & \multicolumn{1}{c|}{0} & 0 & \multicolumn{1}{c|}{\textbf{0}} & \textbf{0} & \multicolumn{1}{c|}{0} & 0 & \multicolumn{1}{c|}{{\color[HTML]{FE0000} \textbf{0}}} & {\color[HTML]{FE0000} \textbf{0}} \\ \cline{2-18} 
\multirow{-3}{*}{\textbf{Rastrigin + noise}} & 80 & \multicolumn{1}{c|}{319} & 68 & \multicolumn{1}{c|}{169} & 55.6 & \multicolumn{1}{c|}{58.1} & 18.3 & \multicolumn{1}{c|}{53} & 18 & \multicolumn{1}{c|}{0} & 0 & \multicolumn{1}{c|}{\textbf{0}} & \textbf{0} & \multicolumn{1}{c|}{0} & 0 & \multicolumn{1}{c|}{{\color[HTML]{FE0000} \textbf{0}}} & {\color[HTML]{FE0000} \textbf{0}} \\ \hline
\end{tabular}
\end{adjustbox}
\label{statistik_100-rastrigin}
\end{table}

\begin{table}[]
 \centering
\caption{Statistical results from 100 runs over the $Sphere$-Function}
\begin{adjustbox}{max width=\paperwidth, angle=90}
\begin{tabular}{|c|c|cc|cc|cc|cc|cc|cc|cc|cc|}
\hline
 &  & \multicolumn{2}{c|}{\textbf{bPSO}} & \multicolumn{2}{c|}{\textbf{bPSO-chi}} & \multicolumn{2}{c|}{\textbf{SPSA--PSO (1)}} & \multicolumn{2}{c|}{\textbf{SPSA--PSO-chi (1)}} & \multicolumn{2}{c|}{\textbf{SPSA--PSO (2)}} & \multicolumn{2}{c|}{\textbf{SPSA--PSO-chi (2)}} & \multicolumn{2}{c|}{\textbf{SPSA--PSO (3)}} & \multicolumn{2}{c|}{\textbf{SPSA--PSO-chi (3)}} \\ \cline{3-18} 
\multirow{-2}{*}{\textbf{Functions}} & \multirow{-2}{*}{\textbf{d}} & \multicolumn{1}{c|}{\textbf{$\mu$}} & \textbf{$\sigma$} & \multicolumn{1}{c|}{\textbf{$\mu$}} & \textbf{$\sigma$} & \multicolumn{1}{c|}{\textbf{$\mu$}} & \textbf{$\sigma$} & \multicolumn{1}{c|}{\textbf{$\mu$}} & \textbf{$\sigma$} & \multicolumn{1}{c|}{\textbf{$\mu$}} & \textbf{$\sigma$} & \multicolumn{1}{c|}{\textbf{$\mu$}} & \textbf{$\sigma$} & \multicolumn{1}{c|}{\textbf{$\mu$}} & \textbf{$\sigma$} & \multicolumn{1}{c|}{\textbf{$\mu$}} & \textbf{$\sigma$} \\ \hline
 & 20 & \multicolumn{1}{c|}{0} & 0 & \multicolumn{1}{c|}{0} & 0 & \multicolumn{1}{c|}{0} & 0 & \multicolumn{1}{c|}{0} & 0 & \multicolumn{1}{c|}{0} & 0 & \multicolumn{1}{c|}{\textbf{0}} & \textbf{0} & \multicolumn{1}{c|}{0} & 0 & \multicolumn{1}{c|}{{\color[HTML]{FE0000} \textbf{0}}} & {\color[HTML]{FE0000} \textbf{0}} \\ \cline{2-18} 
 & 50 & \multicolumn{1}{c|}{0} & 0 & \multicolumn{1}{c|}{0} & 0 & \multicolumn{1}{c|}{0} & 0 & \multicolumn{1}{c|}{0} & 0 & \multicolumn{1}{c|}{0} & 0 & \multicolumn{1}{c|}{\textbf{0}} & \textbf{0} & \multicolumn{1}{c|}{0} & 0 & \multicolumn{1}{c|}{{\color[HTML]{FE0000} \textbf{0}}} & {\color[HTML]{FE0000} \textbf{0}} \\ \cline{2-18} 
\multirow{-3}{*}{\textbf{Sphere}} & 80 & \multicolumn{1}{c|}{0} & 0 & \multicolumn{1}{c|}{0} & 0 & \multicolumn{1}{c|}{0} & 0 & \multicolumn{1}{c|}{0} & 0 & \multicolumn{1}{c|}{0} & 0 & \multicolumn{1}{c|}{\textbf{0}} & \textbf{0} & \multicolumn{1}{c|}{0} & 0 & \multicolumn{1}{c|}{{\color[HTML]{FE0000} \textbf{0}}} & {\color[HTML]{FE0000} \textbf{0}} \\ \hline
 & 20 & \multicolumn{1}{c|}{0} & 0 & \multicolumn{1}{c|}{0} & 0 & \multicolumn{1}{c|}{0} & 0 & \multicolumn{1}{c|}{0} & 0 & \multicolumn{1}{c|}{0} & 0 & \multicolumn{1}{c|}{\textbf{0}} & \textbf{0} & \multicolumn{1}{c|}{0} & 0 & \multicolumn{1}{c|}{{\color[HTML]{FE0000} \textbf{0}}} & {\color[HTML]{FE0000} \textbf{0}} \\ \cline{2-18} 
 & 50 & \multicolumn{1}{c|}{0} & 0 & \multicolumn{1}{c|}{0} & 0& \multicolumn{1}{c|}{0} & 0 & \multicolumn{1}{c|}{0} & 0 & \multicolumn{1}{c|}{0} & 0 & \multicolumn{1}{c|}{\textbf{0}} & \textbf{0} & \multicolumn{1}{c|}{0} & 0 & \multicolumn{1}{c|}{{\color[HTML]{FE0000} \textbf{0}}} & {\color[HTML]{FE0000} \textbf{0}} \\ \cline{2-18} 
\multirow{-3}{*}{\textbf{Sphere + noise}} & 80 & \multicolumn{1}{c|}{0} & 0 & \multicolumn{1}{c|}{0} & 0 & \multicolumn{1}{c|}{0} & 0 & \multicolumn{1}{c|}{0} & 0 & \multicolumn{1}{c|}{0} & 0 & \multicolumn{1}{c|}{\textbf{0}} & \textbf{0} & \multicolumn{1}{c|}{0} & 0 & \multicolumn{1}{c|}{{\color[HTML]{FE0000} \textbf{0}}} & {\color[HTML]{FE0000} \textbf{0}} \\ \hline
\end{tabular}
\end{adjustbox}
\label{statistik_100-sphere}
\end{table}

\section{Conclusion}

In this paper, we have investigated three methods to improve the convergence of the PSO algorithm. The first and second methods (based on the stochastic PSO algorithm proposed in \cite{Kiranyaz2009StochasticAD}) mainly focused on improving the poor updating of the \textit{gbest} particle. This can be a severe problem, which may cause premature convergence to local optima since $gbest$ as the common term in the update equation of all particles, is the primary guide of the swarm.  The third approach focused on the exploitation (the ability to refine search points that are already good) of the particle in the swarm. Those approaches are tested for three multidimensional non-linear functions with noise and the experimental results demonstrated that they achieved a better performance over all functions regardless of the dimension, modality, etc. Especially if the setting of the critical parameters, $c_1$, $c_2$, $a$, and $c$ is appropriate, a significant performance gain can be achieved by SPSA--PSO. The ability to deal with noise was also here investigated. The experimental results shown that the proposed algorithms achieved a good performance over all functions. But the complexity of SPSA--PSO is not negligible. SPSA--PSO needed more run time than the bPSO. It was shown that the second approach achieved a superior performance in all aspects (dimension, modality, complexity, run time).
%It will then be tested on practical problems to investigate it in comparison to the bPSO.

%\marginpar{Many references were incomplete and simply cited in a way so that nobody has a chance to ever find those documents. On the other hand, some of those documents even provide exactly how they are to be cited on the first page or have downloadable bibtex files.}

\bibliographystyle{IEEEtran}
\bibliography{Article}

\begin{thebibliography}{10}
\providecommand{\url}[1]{#1}
\csname url@samestyle\endcsname
\providecommand{\newblock}{\relax}
\providecommand{\bibinfo}[2]{#2}
\providecommand{\BIBentrySTDinterwordspacing}{\spaceskip=0pt\relax}
\providecommand{\BIBentryALTinterwordstretchfactor}{4}
\providecommand{\BIBentryALTinterwordspacing}{\spaceskip=\fontdimen2\font plus
\BIBentryALTinterwordstretchfactor\fontdimen3\font minus
  \fontdimen4\font\relax}
\providecommand{\BIBforeignlanguage}[2]{{%
\expandafter\ifx\csname l@#1\endcsname\relax
\typeout{** WARNING: IEEEtran.bst: No hyphenation pattern has been}%
\typeout{** loaded for the language `#1'. Using the pattern for}%
\typeout{** the default language instead.}%
\else
\language=\csname l@#1\endcsname
\fi
#2}}
\providecommand{\BIBdecl}{\relax}
\BIBdecl

\bibitem{488968}
J.~Kennedy and R.~Eberhart, ``Particle swarm optimization,'' in
  \emph{Proceedings of ICNN'95 - International Conference on Neural Networks},
  vol.~4, 1995, pp. 1942--1948 vol. 4.

\bibitem{1282114}
J.~Robinson and Y.~Rahmat-Samii, ``Particle swarm optimization in
  electromagnetics,'' \emph{IEEE Transactions on Antennas and Propagation},
  vol.~52, no.~2, pp. 397--407, 2004.

\bibitem{OURIQUE20021783}
\BIBentryALTinterwordspacing
C.~O. Ourique, E.~C. Biscaia, and J.~C. Pinto, ``The use of particle swarm
  optimization for dynamical analysis in chemical processes,'' \emph{Computers
  \& Chemical Engineering}, vol.~26, no.~12, pp. 1783--1793, 2002. [Online].
  Available:
  \url{https://www.sciencedirect.com/science/article/pii/S0098135402001539}
\BIBentrySTDinterwordspacing

\bibitem{PRATA20093953}
\BIBentryALTinterwordspacing
D.~M. Prata, M.~Schwaab, E.~L. Lima, and J.~C. Pinto, ``Nonlinear dynamic data
  reconciliation and parameter estimation through particle swarm optimization:
  Application for an industrial polypropylene reactor,'' \emph{Chemical
  Engineering Science}, vol.~64, no.~18, pp. 3953--3967, 2009. [Online].
  Available:
  \url{https://www.sciencedirect.com/science/article/pii/S0009250909003480}
\BIBentrySTDinterwordspacing

\bibitem{TANG20091391}
\BIBentryALTinterwordspacing
Y.~Tang and X.~Guan, ``Parameter estimation for time-delay chaotic system by
  particle swarm optimization,'' \emph{Chaos, Solitons \& Fractals}, vol.~40,
  no.~3, pp. 1391--1398, 2009. [Online]. Available:
  \url{https://www.sciencedirect.com/science/article/pii/S0960077907007576}
\BIBentrySTDinterwordspacing

\bibitem{9254880}
B.~Arabsalmanabadi, N.~Tashakor, S.~Goetz, and K.~Al-Haddad, ``Li-ion battery
  models and a simplified online technique to identify parameters of electric
  equivalent circuit model for ev applications,'' in \emph{IECON 2020 The 46th
  Annual Conference of the IEEE Industrial Electronics Society}, 2020, pp.
  4164--4169.

\bibitem{9589293}
B.~Arabsalmanabadi, N.~Tashakor, Y.~Zhang, K.~Al-Haddad, and S.~Goetz,
  ``Parameter estimation of batteries in mmcs with parallel connectivity using
  pso,'' in \emph{IECON 2021 – 47th Annual Conference of the IEEE Industrial
  Electronics Society}, 2021, pp. 1--6.

\bibitem{6347004}
S.~Goetz, N.~Truong, M.~Gerhofer, A.~Peterchev, H.-G. Herzog, and T.~Weyh,
  ``Optimization of magnetic neurostimulation waveforms for minimum power
  loss,'' in \emph{2012 Annual International Conference of the IEEE Engineering
  in Medicine and Biology Society}, 2012, pp. 4652--4655.

\bibitem{10.1371/journal.pone.0055771}
\BIBentryALTinterwordspacing
S.~M. Goetz, C.~N. Truong, M.~G. Gerhofer, A.~V. Peterchev, H.-G. Herzog, and
  T.~Weyh, ``Analysis and optimization of pulse dynamics for magnetic
  stimulation,'' \emph{PLOS ONE}, vol.~8, no.~3, pp. 1--12, 03 2013. [Online].
  Available: \url{https://doi.org/10.1371/journal.pone.0055771}
\BIBentrySTDinterwordspacing

\bibitem{Barker1975SociobiologyTN}
E.~Barker and E.~O. Wilson, ``Sociobiology: The new synthesis,'' \emph{British
  Journal of Sociology}, vol.~26, p. 501, 1975.

\bibitem{Bck1993AnOO}
T.~B{\"a}ck and H.-P. Schwefel, ``An overview of evolutionary algorithms for
  parameter optimization,'' \emph{Evolutionary Computation}, vol.~1, pp. 1--23,
  1993.

\bibitem{Goldberg1988GeneticAI}
D.~E. Goldberg, ``Genetic algorithms in search optimization and machine
  learning,'' 1988.

\bibitem{Koza1993GeneticP}
J.~R. Koza, ``Genetic programming - on the programming of computers by means of
  natural selection,'' in \emph{Complex adaptive systems}, 1993.

\bibitem{Bck1995EVOLUTIONARYAF}
T.~B{\"a}ck and F.~Kursawe, ``Evolutionary algorithms for fuzzy logic: A brief
  overview,'' 1995.

\bibitem{Zaki2010AdvancesIK}
M.~J. Zaki, J.~X. Yu, B.~Ravindran, and V.~Pudi, ``Advances in knowledge
  discovery and data mining, 14th pacific-asia conference, pakdd 2010,
  hyderabad, india, june 21-24, 2010. proceedings. part i,'' in \emph{PAKDD},
  2010.

\bibitem{Riget2002ADP}
J.~Riget and J.~Vesterstr{\o}m, ``A diversity-guided particle swarm optimizer
  -- the arpso,'' in \emph{EVALife Technical Report no. 2002-02}, 2002.

\bibitem{Bergh2002AnAO}
F.~van~den Bergh and A.~P. Engelbrecht, ``An analysis of particle swarm
  optimizers,'' 2002.

\bibitem{Helwig2010ParticleSF}
S.~Helwig, ``Particle swarms for constrained optimization,'' 2010.

\bibitem{Seo2006MultimodalFO}
J.~Seo, C.-H. Im, C.~G. Heo, J.-K. Kim, H.-K. Jung, and C.-G. Lee, ``Multimodal
  function optimization based on particle swarm optimization,'' \emph{IEEE
  Transactions on Magnetics}, vol.~42, pp. 1095--1098, 2006.

\bibitem{Chen2010AnIC}
D.~bao Chen, C.~Zhao, and H.~Zhang, ``An improved cooperative particle swarm
  optimization and its application,'' \emph{Neural Computing and Applications},
  vol.~20, pp. 171--182, 2010.

\bibitem{Eberhart2000ComparingIW}
R.~C. Eberhart and Y.~Shi, ``Comparing inertia weights and constriction factors
  in particle swarm optimization,'' \emph{Proceedings of the 2000 Congress on
  Evolutionary Computation. CEC00 (Cat. No.00TH8512)}, vol.~1, pp. 84--88,
  2000.

\bibitem{Clerc2002ThePS}
M.~Clerc and J.~Kennedy, ``The particle swarm - explosion, stability, and
  convergence in a multidimensional complex space,'' \emph{IEEE Trans. Evol.
  Comput.}, vol.~6, pp. 58--73, 2002.

\bibitem{Zyl2015ASM}
E.~T. van Zyl and A.~P. Engelbrecht, ``A subspace-based method for pso
  initialization,'' \emph{2015 IEEE Symposium Series on Computational
  Intelligence}, pp. 226--233, 2015.

\bibitem{Richards2004ChoosingAS}
M.~Richards and D.~Ventura, ``Choosing a starting configuration for particle
  swarm optimization,'' \emph{2004 IEEE International Joint Conference on
  Neural Networks (IEEE Cat. No.04CH37541)}, vol.~3, pp. 2309--2312 vol.3,
  2004.

\bibitem{Engelbrecht2005FundamentalsOC}
A.~P. Engelbrecht, ``Fundamentals of computational swarm intelligence,'' 2005.

\bibitem{Spall1998ImplementationOT}
J.~C. Spall, ``Implementation of the simultaneous perturbation algorithm for
  stochastic optimization,'' \emph{IEEE Transactions on Aerospace and
  Electronic Systems}, vol.~34, pp. 817--823, 1998.

\bibitem{Kiranyaz2009StochasticAD}
S.~Kiranyaz, T.~Ince, and M.~Gabbouj, ``Stochastic approximation driven
  particle swarm optimization,'' \emph{2009 International Conference on
  Innovations in Information Technology (IIT)}, pp. 40--44, 2009.

\bibitem{Kiranyaz2010FractionalPS}
S.~Kiranyaz, T.~Ince, E.~A. Yildirim, and M.~Gabbouj, ``Fractional particle
  swarm optimization in multidimensional search space,'' \emph{IEEE
  Transactions on Systems, Man, and Cybernetics, Part B (Cybernetics)},
  vol.~40, pp. 298--319, 2010.

\bibitem{Angeline1998EvolutionaryOV}
P.~J. Angeline, ``Evolutionary optimization versus particle swarm optimization:
  Philosophy and performance differences,'' in \emph{Evolutionary Programming},
  1998.

\bibitem{Esquivel2003OnTU}
S.~C. Esquivel and C.~A.~C. Coello, ``On the use of particle swarm optimization
  with multimodal functions,'' \emph{The 2003 Congress on Evolutionary
  Computation, 2003. CEC '03.}, vol.~2, pp. 1130--1136 Vol.2, 2003.

\bibitem{Shi1998AMP}
Y.~Shi and R.~C. Eberhart, ``A modified particle swarm optimizer,'' \emph{1998
  IEEE International Conference on Evolutionary Computation Proceedings. IEEE
  World Congress on Computational Intelligence (Cat. No.98TH8360)}, pp. 69--73,
  1998.

\end{thebibliography}

\end{document}